\documentclass[aps,prb,twocolumn,groupedaddress]{revtex4-2}
\usepackage{amsmath}
\usepackage{amssymb}
\usepackage{amsfonts}
\usepackage{graphicx}
\usepackage{bm}
\usepackage{tikz}
\usepackage{adjustbox}
\usepackage{physics}
\usepackage{comment}
\usepackage[colorlinks=true, linkcolor=blue, citecolor=blue, urlcolor=blue]{hyperref}
\begin{document}

\title{Gaussian fixed lines of $S=1/2$ XXZ chain with next-nearest neighbor interaction and $sl_2$ loop algebra}
\author{Daiki Yomatsu and Kiyohide Nomura}
\affiliation{Department of Physics, Kyushu University, Fukuoka, Japan, 819-0395}

\date{\today}

\begin{abstract}
Spin systems are important to understand various physical properties in quantum many-body systems. 
We numerically study the Gaussian fixed lines (GFLs) of the $S=1/2$ XXZ chain with next-nearest neighbor (NNN) interaction in the XY phase. 
The GFLs are the set of points where the coefficient of the umklapp scattering vanishes. 
We show that the GFLs pass through the ``special points'', which are defined as the points where the $sl_2$ loop algebra symmetry governs in low-energy physics.
In addition, we have discussed the Tomonaga-Luttinger parameter $K$, and the metamagnetism influences the shape of the GFLs. 
\end{abstract}

\maketitle
\section{Introduction}
%%%%%%%%%%%%%%追加した導入
In condensed matter physics, the umklapp scattering (or umklapp process) plays an important role in transport phenomena~\cite{Ashcroft-Mermin:1976aa} and nonlinear responses~\cite{Tanikawa:2021aa,Tanikawa:2021ab}. 
But in some cases, the umklapp coupling disappears. 
For one-dimensional quantum spin model, the Gaussian fixed lines (GFLs) mean the disappearance of the umklapp coupling.%この文は後で変更or補足が必要かも
%%%%%%%%%%%%%%
\par
In this paper, the GFLs are the set of points in the quantum sine-Gordon model, which appears as the effective theory of one-dimensional quantum many-body systems, where the coefficient of the cosine term becomes zero. 
This cosine term corresponds to the umklapp scattering in fermion models on lattice~\cite{Giamarchi:2003aa}. 
\par
We treat the $S=1/2$ XXZ chain with next-nearest neighbor (NNN) interaction. 
The Hamiltonian is given by
\begin{align}
      H(\Delta,\alpha) &{} = \sum_{i=1}^{L} \left( S_i^x S_{i+1}^x + S_i^y S_{i+1}^y + \Delta S_i^z S_{i+1}^z \right) \notag \\
      &{} + \alpha \sum_{i=1}^{L} \left( S_i^x S_{i+2}^x + S_i^y S_{i+2}^y + \Delta S_i^z S_{i+2}^z \right), \label{eq: hamiltonian}
\end{align}
where we assume the periodic boundary condition (PBC) $S_{L+i}^a=S_i^a$, and $L$ is the number of spins ($L$ even). 
The coupling $\alpha$ is the next-nearest neighbor interaction, and $\Delta$ is the anisotropy for the $z$ component. 
Under the PBC, the eigenstates are labeled by the magnetization $m=\sum_i S^z_i$, the wave number $k=\frac{2\pi}{L}n$ ($n\in\mathbb{Z}$), the (bond-centered) parity $P=\pm1$, and the time-reversal symmetry $T=\pm1$. 
\par
Some specific cases of the Hamiltonian~(\ref{eq: hamiltonian}) have been investigated a lot.
For example, the case of $\alpha=0$ is known to one of the quantum integrable systems by the Bethe ansatz~\cite{Bethe:1931aa,Des-Cloizeaux:1966aa}. 
More especially, at $\Delta=0,\alpha=0$, the Hamiltonian~(\ref{eq: hamiltonian}) is called the isotropic XY model, which is equivalent to a free fermion model, and is solvable without Bethe ansatz~\cite{Lieb:1961aa}. 
For $-1 < \Delta < 1,\alpha=0$ region, the ground state is the XY state, which is characterized by a gapless exitation and the power-law decay of the spin correlation functions. 
Then, the isotropic ($\Delta=1$) and the NNN interaction ($\alpha\ne0$) case, namely the $J_1-J_2$ model, has been studied~\cite{Tonegawa1987:aa,Okamoto1992,Farnell1994:aa}.
The case of $\Delta=1, \alpha=1/2$ is called the Majumdar-Ghosh (MG) model~\cite{Majumdar:1969aa,Majumdar:1969ab}.
The ground state of MG model is the purely dimer state (also known as the valence bond solid)
\begin{align*}
      \ket{\Psi_1} &=[1,2]\otimes[3,4]\otimes\cdots\otimes[L-1,L]\\
      \ket{\Psi_2} &=[2,3]\otimes[4,5]\otimes\cdots\otimes[L,1],
\end{align*}
where $[i,j]$ is the singlet state of the $i$-th and $j$-th spins~\cite{VANDENBROEK1980261}.
\par
\begin{figure}[tbp]
      \centering
      \includegraphics[width=\linewidth]{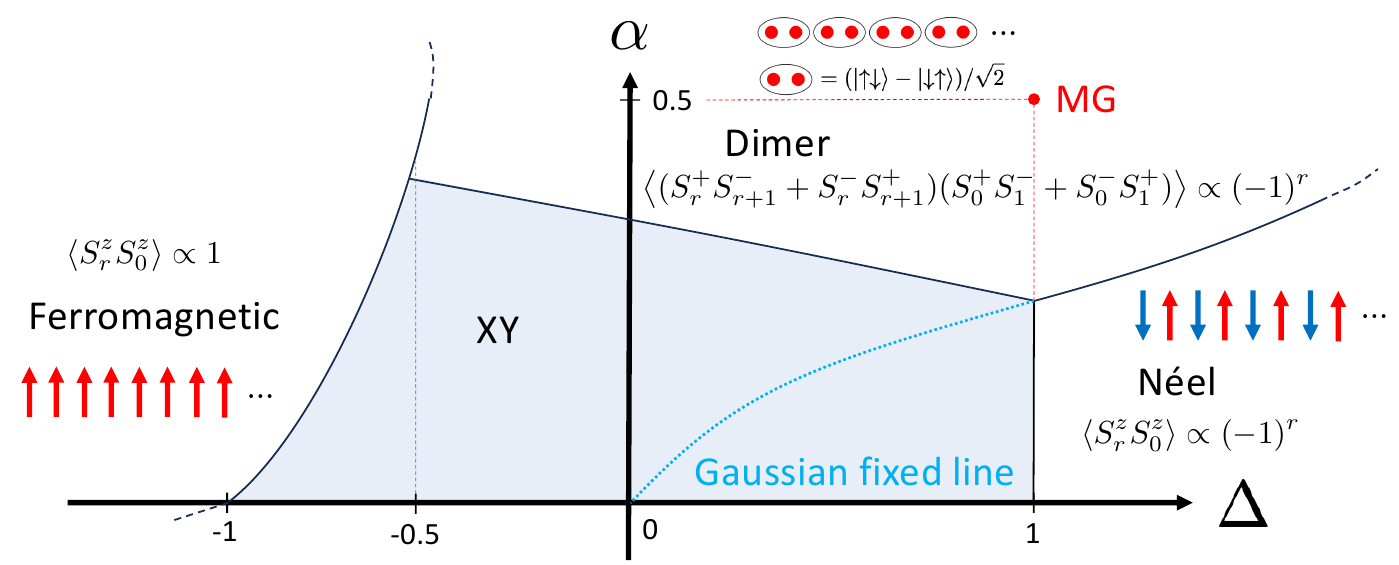}
      \caption{(Color online)
            Schematic phase diagram of the Hamiltonian (\ref{eq: hamiltonian}). 
            It is divided into four phases: FM, N\'{e}el, dimer and XY phases. 
            The XY phase is shaded region. 
            }
      \label{fig:phase_diagram}
\end{figure}

The phase diagram of the Hamiltonian~(\ref{eq: hamiltonian}) has partially obtained~\cite{Nomura:1993aa,Hirata:2000aa}. 
Their results are schematically shown in Fig.~\ref{fig:phase_diagram}. 
The phase diagram is divided into four phases: ferromagnetic (FM), N\'{e}el, dimer and XY phases. 
The ground state of the FM phase is fully polarized state ($m=\pm L/2$) and its energy is given by $E_{\rm FM}=\frac{L}{4}(1+\alpha)\Delta$. 
The dimer and N\'{e}el phases are characterized by the exitation gap, the twofold ground state and the long-range dimer and N\'{e}el order. 
The XY phase is a critical phase and also called the Tomonaga-Luttinger liquid (TLL) phase~\cite{Furukawa:2010aa,Nakamura:1997aa}, the quantum spin liquid phase~\cite{Dutton:2012aa} or the spin fluid phase~\cite{Nomura:1994aa,Hirata:2000aa}. 
Its critical behaviors are investigated using the sine-Gordon model~\cite{Nomura:1994aa}, and also studied the GFLs for $\Delta\ge0$, which is shown in Fig.~\ref{fig:phase_diagram}.

\par
On the GFLs, this umklapp term vanishes, and the low-energy effective theory is described by the Gaussian model, which is a free boson field theory known as a conformal field theory (CFT) with central charge $c=1$. 
The CFT provides a powerful framework for determining the scaling behavior and critical exponents of one-dimensional quantum critical systems, and the GFLs are one of its typical realizations.
\par
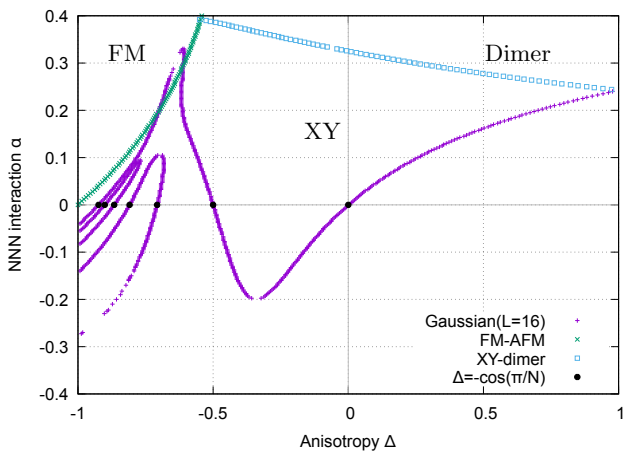
\begin{figure}[htbp]
      \centering
      \begin{tikzpicture}
            \node[anchor=south west,inner sep=0] (duck) at (0,0) {
            \includegraphics[width=\linewidth]{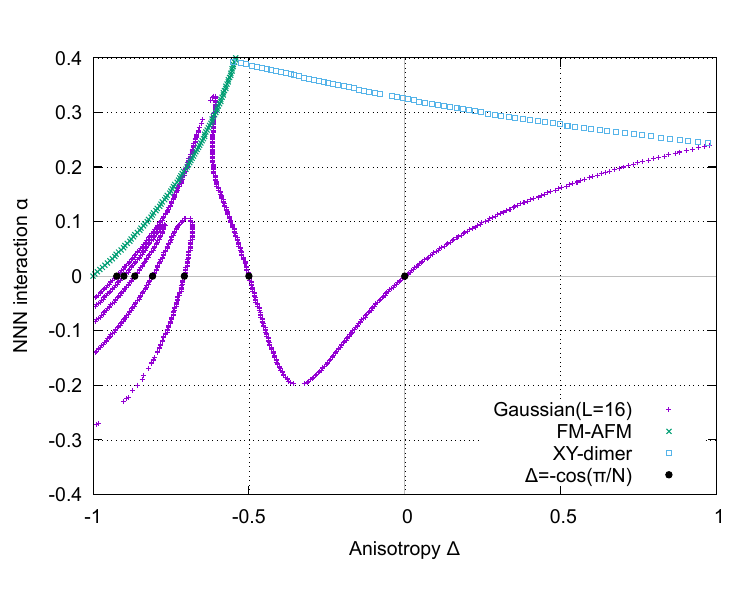}
            };
            \begin{scope}[x={(duck.south east)},y={(duck.north west)}]
                  \node[anchor=south] at (0.2,0.8) {FM};
                  \node[anchor=south] at (0.5,0.65) {XY};
                  \node[anchor=south] at (0.8,0.8) {Dimer};
            \end{scope}
      \end{tikzpicture}
      \caption{(Color online) The Gaussian fixed lines for $L=16$. }
      \label{fig:GFL_L16}
\end{figure}

In this paper, we study the GFLs of the $S=1/2$ XXZ chain with NNN interaction using the level spectroscopy for wider range than previous study \cite{Nomura:1994aa}. 
Our main result is shown in Fig.~\ref{fig:GFL_L16}. 
The GFLs for $L=16$ are shown in purple ``+''.
The FM-XY and XY-dimer transition lines are shown in green ``$\times$'' and blue ``$\square$'', respectively.
The black dots are the ``special points'', which are explicity given by $\Delta=-\cos(\pi/N)$ with $N\in\{ 2,3,4,5,6,7,8\}$ for $L=16$ case. 
We shall provide a general definition of the special points later (see Sec.~\ref{sec: level crossing and special points}, Eq.~(\ref{eq: special points})).
We find that the GFLs pass through the special points, which reflects the structure of the $sl_2$ loop algebra.
This further suggests that some of the GFLs are connected due to underlying physical mechanisms.

\par
Our paper is organized as follows: 
In Sec.~\ref{sec: special symmetry of XXZ chain and level crossings}, we review the quantum group $U_q(sl_2)$ and the $sl_2$ loop algebra as special symmetry of the Hamiltonian (\ref{eq: hamiltonian}) of $\alpha=0$ case and we introduce the ``special points''. Then we discuss about level crossings for the Hamiltonian (\ref{eq: hamiltonian}) with $\Delta=-1/2$. In Sec.~\ref{sec: effective field theory}, we review the quantum sine-Gordon model as the low-energy effective field theory of the Hamiltonian (\ref{eq: hamiltonian}) and introduce the GFLs.
In Sec.~\ref{sec: GFL}, we show the numerical results of the GFLs using the level spectroscopy method. Then we discuss how the special points influence the GFLs and the finite size scaling of the GFLs.
In Sec.~\ref{sec: discussion}, we discuss the shape of GFLs using the Tomonaga-Luttinger (TL) parameter and the metamagnetism. 
Finally, we conclude in Sec.~\ref{sec: conclusion}.

\section{Special symmetries of XXZ chain and level crossings}\label{sec: special symmetry of XXZ chain and level crossings}
Deguchi {\it et al.}~\cite{Deguchi:2001aa} found that the system~(\ref{eq: hamiltonian}) possesses quantum group $U_q(sl_2)$ and $sl_2$ loop algebra symmetry for $-1<\Delta<1$ and $\alpha=0$. 
\par
In this section, we review the quantum group $U_q(sl_2)$ and $sl_2$ loop algebra symmetries of the Hamiltonian~(\ref{eq: hamiltonian}) with $\alpha=0$ (denoted as $H(\Delta,0)$) and discuss the associated level crossings. 
Then we also review the level crossings for $\alpha\ne0$ case, which was discussed by Gerhardt {\it et al.}~\cite{Gerhardt:1998aa} in the final part of this section. 

\subsection{Quantum group symmetry}\label{sec: quantum group symmetry}
The quantum group $U_q(sl_2)$ has generators $q^{\pm S_q^z}, S_q^{\pm}$, and is defined by the relations
\begin{align}
      q^{S_q^z}q^{-S_q^z}&=1,\notag\\
      q^{S_q^z}S_q^{\pm}q^{-S_q^z}&=q^{\pm 1}S_q^{\pm},\label{eq: QG rel.}\\
      [S_q^+,S_q^-]&=\frac{q^{2S_q^z}-q^{-2S_q^z}}{q-q^{-1}},\notag
\end{align}
where $q\in\mathbb{C}$ is the parameter of $U_q(sl_2)$, and satisfies $q^{2N}=1\,(N\in \mathbb{N},q\neq \pm 1)$. Next, we define the operators $S_q^{\pm(N)}$ as
\begin{align}
      S_q^{\pm(N)}:=\lim_{q^{2N}\to 1}(S_q^{\pm})^N/[N]! ,
\end{align}
where $[n]!=\prod_{k=1}^n [k], [n]=\frac{q^n-q^{-n}}{q-q^{-1}}$. The operator $S_q^{\pm(N)}$ can be expressed using the Pauli matrices $\sigma^{\pm}_i,\sigma^{z}_j$ as
\begin{align}
      S_q^{\pm(N)}=\sum\,
            &q^{\frac{N}{2}\sigma^z_1}\otimes \cdots\otimes q^{\frac{N}{2}\sigma^z_{j_1-1}}\notag\\
            &\otimes\sigma_{j_1}^{\pm}\otimes q^{\frac{N-2}{2}\sigma^z_{j_1+1}}
            \otimes \cdots\otimes q^{\frac{N-2}{2}\sigma^z_{j_2-1}}\notag\\
            &\otimes\sigma_{j_2}^{\pm}\otimes q^{\frac{N-4}{2}\sigma^z_{j_2+1}}
            \otimes\cdots\otimes q^{-\frac{N-2}{2}\sigma^z_{j_N-1}}\notag\\
            &\otimes\sigma_{j_N}^{\pm}\otimes q^{-\frac{N}{2}\sigma^z_{j_N+1}}\otimes\cdots\otimes q^{-\frac{N}{2}\sigma^z_{L-1}}\notag\\
            &\otimes q^{-\frac{N}{2}\sigma^z_L},\label{eq: rep. of S_q^N}
\end{align}
where summation is taken over all $N$-combinations of the sites $1\le j_1<j_2<\cdots<j_N\le L$~\cite{Pasquier:1990aa}. When the parameter $q$ satisfies $q^{2N}=1$ and the anisotropy $\Delta$ is given by
\begin{align}
      \Delta = \frac{1}{2}(q+q^{-1}).
\end{align}
At this point, in $m\equiv0 \pmod N$ sector, 
the Hamiltonian~(\ref{eq: hamiltonian}) with $\alpha=0$ commutes with the operators $S_q^{\pm(N)}$ :
\begin{align}\label{eq: XXZ vs sl_2 loop alg.}
      [S_q^{\pm(N)},H(\Delta,\alpha=0)]=0.
\end{align}
\par
When the operator $S_q^{\pm(N)}$ acts on a quantum state, it increases (or decreases) the magnetization by $N$, as seen from Eq.~(\ref{eq: rep. of S_q^N}). 
This fact restricts the possible values of $N$ to $N=2,3,\cdots,L/2$. Thus, the points, where $U_q(sl_2)$ symmetry emerges, depend on the system size. 
The points, where $U_q(sl_2)$ symmetry emerges, are 
\begin{align}
      \Delta=\cos\qty(\frac{\pi}{N}n),\quad n\in\{1,\cdots,N-1\}. \label{eq: U_q(sl_2) symmetry emarges}
\end{align}

\subsection{loop algebra symmetry}\label{sec: loop algebra symmetry}
Next, we introduce the $sl_2$ loop algebra. This is defined as $sl(2,\mathbb{C})\otimes\mathbb{C}[z,z^{-1}]$. The elements of the $sl_2$ loop algebra are expressed as $L^a\otimes z^n (z\in\mathbb{C},n\in\mathbb{Z})$ using the operators $L^a$ of $sl(2,\mathbb{C})$, and the commutation relations are defined as
\begin{align}\label{eq: sl_2 loop alg}
      [L^a\otimes z^n,L^b\otimes z^m]=[L^a,L^b]\otimes z^{n+m}.
\end{align}
The Chevalley basis of the $sl_2$ loop algebra can be expressed using the Chevalley basis $e,f,h$ of $sl(2,\mathbb{C})$ as
\begin{align}
      \begin{split}
            &h_0=-h\otimes 1,	\quad e_0=f\otimes z,	\quad f_0=e\otimes z^{-1},\\
            &h_1=h\otimes 1,	\quad e_1=e\otimes 1,	\quad f_1=f\otimes 1,
      \end{split}
\end{align}
and satisfies the following relations
\begin{align}
      &[h_0,h_1]=0,\, [h_i,e_j]=a_{ij}e_j,\notag\\
      &[h_i,f_j]=-a_{ij}f_j,\, [e_i,f_j]=\delta_{ij}h_j, \label{eq: serre rel.}\\
      &[e_i,[e_i,[e_i,e_j]]]=0,\, [f_i,[f_i,[f_i,f_j]]]=0, \, (i\neq j)\notag
\end{align}
where $i,j=0,1$ and $a_{ij}$ are elements of the Cartan matrix defined as
\begin{align}
      \begin{pmatrix}
            a_{00} & a_{01}\\
            a_{10} & a_{11}
      \end{pmatrix}
      =
      \begin{pmatrix}
            2 	& -2\\
            -2 	& 2
      \end{pmatrix}.
\end{align}
Then, using $S_q^{\pm(N)},S_q^z$ and $T_q^{\pm(N)} := S_{q^{-1}}^{\pm(N)}$, we define
\begin{align}
      \begin{split}
      &h_0=-(-q)^NS_q^z/N,\quad e_0=S_q^{+(N)},\quad f_0=S_q^{-(N)}\\
      &h_1=(-q)^NS_q^z/N,\quad e_1=T_q^{-(N)},\quad f_1=T_q^{+(N)}.
      \end{split}
\end{align}
When $q^{2N}=1$, these satisfy the commutation relations~(\ref{eq: serre rel.}) in $m\equiv0 \pmod N$ sector~\cite{Deguchi:2001aa}. 
Thus, $S_q^{\pm(N)},T_q^{\pm(N)},S_q^z$ generate the $sl_2$ loop algebra. From the above, in the Hamiltonian $H(\Delta,0)$ with specific anisotropy $\Delta$ that expressed by Eq.~(\ref{eq: U_q(sl_2) symmetry emarges}), the $sl_2$ loop algebra symmetry emerges.

\subsection{Level crossings and special points}\label{sec: level crossing and special points}
Now, $S_q^{\pm(N)}$ changes the magnetization by $\pm N$, and the commutation relations~(\ref{eq: XXZ vs sl_2 loop alg.}) implies that the eigenstates with $m=0$ and those with $m=\pm N,\pm 2N,\cdots$ cause level crossings (or energy degeneracies). 
\par
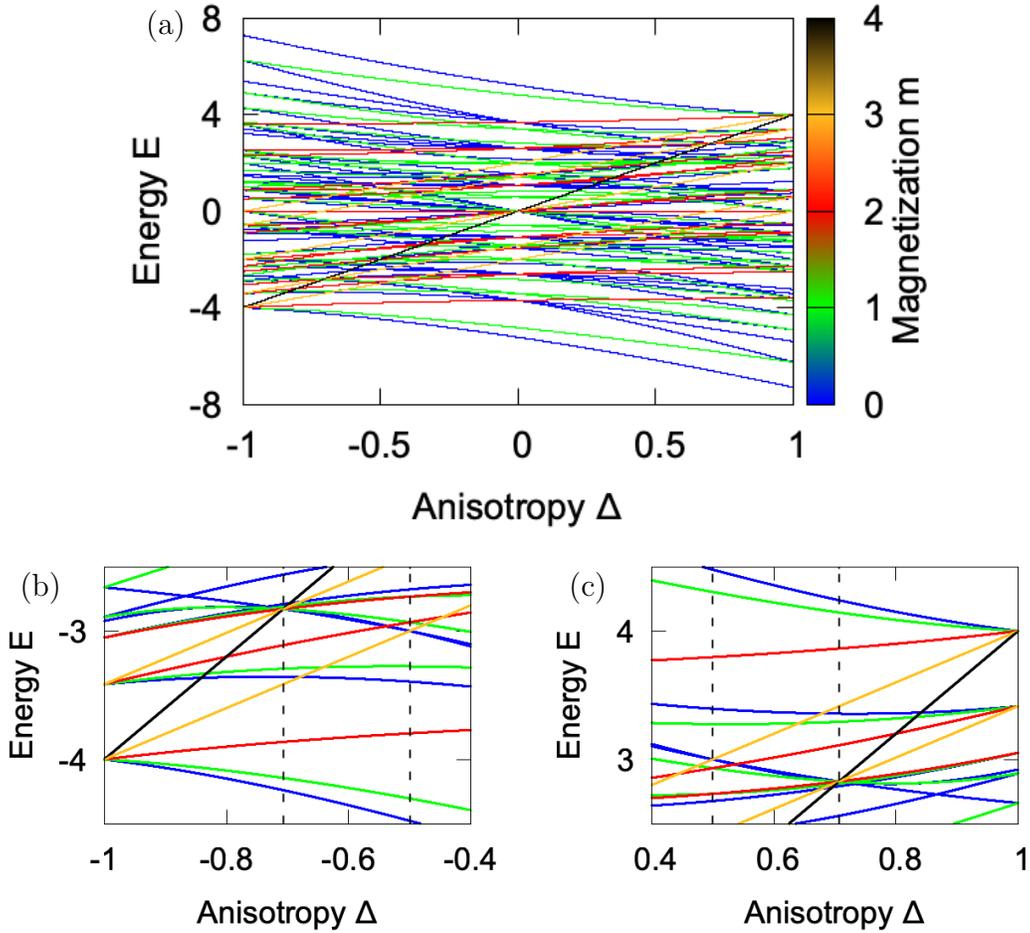
\begin{figure*}[tbp]
      \centering
      \begin{minipage}{0.6\linewidth}
            \centering
            \begin{tikzpicture}
                  \node[anchor=south west,inner sep=0] (duck) at (0,0) {
                  \includegraphics[width=\linewidth]{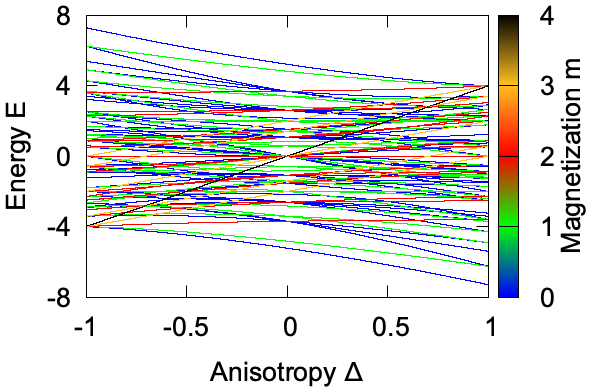}
                  };
                  \begin{scope}[x={(duck.south east)},y={(duck.north west)}]
                        \node[anchor=south] at (0.05,0.9) {\large (a)};
                  \end{scope}
            \end{tikzpicture}
      \end{minipage}
      \begin{minipage}{0.4\linewidth}
            \centering
            \begin{tikzpicture}
                  \node[anchor=south west,inner sep=0] (duck) at (0,0) {
                  \includegraphics[width=\linewidth]{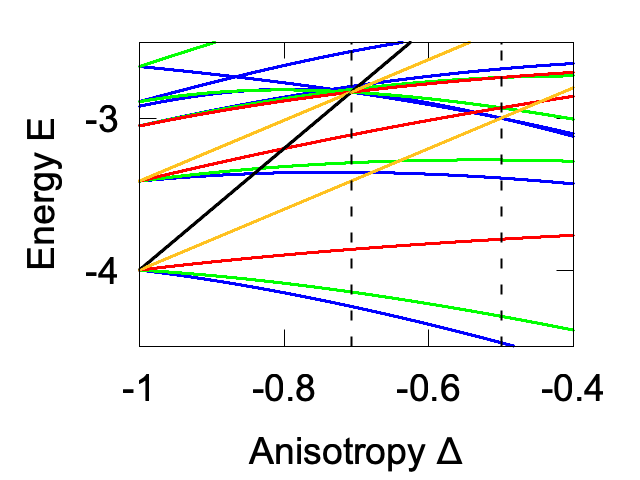}
                  };
                  \begin{scope}[x={(duck.south east)},y={(duck.north west)}]
                        \node[anchor=south] at (0.1 ,0.8) {\large (b)};
                  \end{scope}
            \end{tikzpicture}
      \end{minipage}
      \begin{minipage}{0.4\linewidth}
            \centering
            \begin{tikzpicture}
                  \node[anchor=south west,inner sep=0] (duck) at (0,0) {
                  \includegraphics[width=\linewidth]{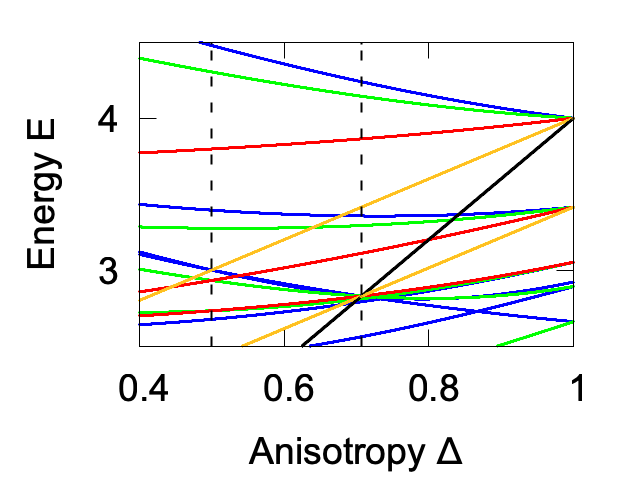}
                  };
                  \begin{scope}[x={(duck.south east)},y={(duck.north west)}]
                        \node[anchor=south] at (0.1 ,0.8) {\large (c)};
                  \end{scope}
            \end{tikzpicture}
      \end{minipage}
      \caption{(Color online) 
            (a) shows the full energy spectrum of the $H_{\rm NN}$ ($L = 8$) as a function of the anisotropy $\Delta$. 
            The color of the lines represents the magnetization of the states.
            (b) and (c) show the lower and higher parts of the energy spectrum, respectively. The black dashed lines indicate $\Delta = -1/\sqrt{2}, -1/2$ in (b) and $\Delta = +1/\sqrt{2}, +1/2$ in (c). 
      }
      \label{fig:energy_spectrum}
\end{figure*}

We calculate the energy spectrum of the Hamiltonian $H(\Delta,0)$ with $L=8$ by numerical diagonalization (Fig.~\ref{fig:energy_spectrum}). It shows level crossings between $m=0$ states and $m=3,4$ states can be observed at $\Delta=\pm 1/2,\pm 1/\sqrt{2}$, respectively. 
In the region $\Delta<0$, eigenstates with high magnetization (e.g., $m=3,4$ in $L=8$ systems) appear as low excitations because of the region  $\Delta<-1$ is the FM phase.
In contrast, in the region $\Delta>0$, it can be seen that eigenstates with high magnetization appear as high excitations. 
Thus, in the region $\Delta>0$, level crossings due to the $sl_2$ loop algebra appear in higher excitations and do not contribute to the low-energy physics. 
\par
From these considerations, it can be understood that the $sl_2$ loop algebra contributes to the low-energy physics when
\begin{align}
      \Delta&{}=\cos\qty(\frac{(N-1)\pi}{N})\notag\\
      &{}=-\cos(\pi/N),\label{eq: special points}
\end{align}
where $N\in\{2,3,4,\cdots,L/2\}$.
According to \cite{Furukawa:2010aa}, they call these points ``special points''. 
We also use the term special points in this paper.
At the special points, it has been reported that certain physical quantities exhibit anomalous behavior.
Recently, Tanikawa {\it et al.}~\cite{Tanikawa:2021ab} reported the non-linear Drude weights converge to a finite value at the special points (they have called it ``exceptional points''), but diverges at other points.

\subsection{Level crossings with NNN interactions}\label{sec: level crossings with NNN interactions}
Gerhardt {\it et al.}~\cite{Gerhardt:1998aa} discussed level crossings for $\Delta=-1/2$ and $\alpha\in\mathbb{R}$ at Section IV in their paper. 
In particular, for $L=6n, n\in\mathbb{N}$, they analytically derived the following rigorous relation:
\begin{align}
      \qty[H \qty(\Delta=-\frac{1}{2},\alpha),S_{+}\qty(p=\pm\frac{2\pi}{3})]\ket{F-}=0, \label{eq: Gerhardt degeneracies}
\end{align}
where $S_{+}(p)=\sum_{l} e^{ipl}S_l^+$ and $\ket{F-}$ denotes the fully polarized state with all spins aligned in the negative $z$ direction.
There are many more level crossings occurring for all values of $m=0,1,\cdots,L/2$. 
Particulally, for $0.5<\alpha<0.6$, they reported that the level crossings appear even at the ground state using numerical calculations. 
\par
In addition, when we apply the longitudinal magnetic field $h$, the level crossings (Eq.~(\ref{eq: Gerhardt degeneracies})) are broken and make equally spaced energy gaps, which are similar to restricted spectrum generating algebra~(RSGA), suggesting a connection to the quantum many-body scar~(QMBS)~\cite{Moudgalya:2020aa}.

\section{Effective field theory}\label{sec: effective field theory}
The $S=1/2$ XXZ chain with NNN interactions can be mapped on the quantum sine-Gordon model as an effective model, through the Jordan-Wigner transformation~\cite{Jordan:1928} and bosonization~\cite{Haldane:1982aa}. 
The Hamiltonian of the quantum sine-Gordon (SG) model is given by
\begin{align}
      H_{\rm SG} &= \frac{v}{2\pi}\int dx\, \qty[ K(\pi\Pi)^2 + \frac{1}{K}(\partial_x \phi)^2 ] \notag \\
      &\quad + \frac{v y_{\phi}}{2\pi a^2} \int dx\, \cos(\sqrt{8}\phi), \label{sine-Gordon}
\end{align}
where $a$, and $v$ represent the lattice spacing, and the spin-wave velocity, respectively.
Additionally, $K$ and $y_{\phi}$ are coupling constants determined by the parameters $\Delta$ and $\alpha$ in the Hamiltonian~(\ref{eq: hamiltonian}).
The cosine term in the SG model arises from the umklapp scattering of the fermion many body system.
Here, $\Pi$ is the momentum density field conjugate to $\phi$, satisfying the canonical commutation relation $[\phi(x), \Pi(x')] = i\delta(x-x')$.
We introduce $\theta$, the field dual to $\phi$, defined by
\begin{align}
      \partial_x \theta = \pi \Pi.
\end{align}
We assume periodic boundary conditions in the spatial direction, $\phi(x) = \phi(x + L)$, which renders the SG model a field theory on a cylinder.
Furthermore, the two fields $\phi$ and $\theta$ are compactified on a circle of radius $R = 1/\sqrt{2}$:
\begin{align}
      \phi \equiv \phi + \frac{2\pi}{\sqrt{2}}, \quad \theta \equiv \theta + \frac{2\pi}{\sqrt{2}}.
\end{align}
Through the bosonization procedure, the spin operators are expressed as
\begin{align}
      \begin{split}
            S^z_{x/a} &\simeq \frac{a}{\sqrt{2}\pi} \partial_x \phi 
                  + \frac{1}{\pi} e^{i\pi x/a} \cos\sqrt{2} \phi, \\
            S^{\pm} &\simeq e^{\mp i\sqrt{2} \theta}.
      \end{split}
      \label{correspondence between spin and boson}
\end{align}
\par
When $y_{\phi}\neq0$, the $U(1)$ symmetry of the field $\phi$ is broken, but the $U(1)$ symmetry of the field $\theta$ is preserved. 
This represents the $U(1)$ symmetry of the model~(\ref{eq: hamiltonian}) related to the conservation of magnetization $m$. 
The renormalization group (RG) equations for the SG model~(\ref{sine-Gordon}) are given by
\begin{align}
      \dv{y_{\phi}}{\ell}&=-y_{\phi}(\ell)y_0(\ell)\\
      \dv{y_0}{\ell}&=-y_{\phi}^2(\ell)
\end{align}
where $K=1+y_0/2$ and $\ell=\ln L$~\cite{Amit:1980aa}. 
This reveals the existence of three critical lines: $y_{\phi}=0$ corresponds to the GFLs, and $y_{\phi}=\pm y_0\, (y_0>0)$ correspond to the BKT transition lines. 
All points in the region between the BKT transition lines renormalize to the GFLs, which correspond to the XY phase. 
Then, the coupling constants $y_{\phi}, K$ change according to the RG equations, and within the XY phase, they flow to the renormalized values $y_{\phi}=0,K^*$ (stable fixed point).

\subsection{Gaussian model}\label{sec: Gaussian model}
On the line where $y_{\phi}=0$, the theory in Eq.~(\ref{sine-Gordon}) reduces to the Gaussian model (which is equivalent to TLL in $(1+1)$-dimensional spacetime), and the Hamiltonian is given by
\begin{align}\label{eq: hamiltonian gaussian}
      H_{\rm G}=\frac{v}{2\pi}\int dx\, \qty[K(\pi\Pi)^2+\frac{1}{K}(\partial \phi)^2],
\end{align}
where $K$ is the Tomonaga-Luttinger (TL) parameter. 
The condition $y_{\phi} = 0$ in Eq.~(\ref{sine-Gordon}) means that the umklapp term vanishes. 
\par
In the case of $\alpha=0$, the exact solution of the Hamiltonian~(\ref{eq: hamiltonian}) using the Bethe ansatz provides the TL parameter as %~\cite{Luther-Peschel:1975aa}%Luther and Peschelの論文
\begin{align}\label{eq: K exact}
      K^*=\frac{\pi}{\pi-\arccos\Delta}.
\end{align}
This expression indicates that $K^*$ decreases monotonically from $K^*=\infty$ at $\Delta=-1$ to $K^*=1$ at $\Delta=1$. 
Then, the special points (labeled by $N$) defined by Eq.~(\ref{eq: special points}) correspond to $K^*=N$.
\par
The Gaussian model is well-known as a $c=1$ CFT, and the vertex operators are given by the following expression~\cite{GINSPARG1988153}:
\begin{align}\label{eq: Vertex operator}
      \mathcal{O}_{m,n}=:e^{i\sqrt{2}n\phi}e^{i\sqrt{2}m\theta}: .
\end{align}
Here, $:AB:$ denotes normal ordering. The integers $m$ and $n$ correspond to the magnetization and winding number of the compactified boson $\phi$, respectively. 
The scaling dimension $x_{m,n}$ and conformal spin $s_{m,n}$ are given by
\begin{align}\label{eq: scaling dimension with gaussian}
      x_{m,n}=\frac{1}{2}\qty(\frac{m^2}{K}+Kn^2),\quad s_{m,n}=mn.
\end{align}
The excitation gap $\Delta E_{m,n}:=E_{m,n}-E_{\rm 0}$ is related to the scaling dimension as
\begin{align}\label{eq: energy gap}
      \Delta E_{m,n}=\frac{2\pi v}{L}x_{m,n},
\end{align}
where $E_{m,n}$ is the eigenenergy of $\mathcal{O}_{m,n}$ and $E_{\rm 0}$ is the ground state energy.
\par
Using the expression~(\ref{correspondence between spin and boson}), several correlation functions on the GFLs can be obtained as follows~\cite{Nomura:1994aa}
\begin{align}
      &\ev*{S^+_rS^-_0}\propto e^{i\pi r}  \ev{e^{-i\sqrt{2}\theta(r)}\cdot e^{i\sqrt{2}\theta(0)}}\notag\\
            &\quad\quad\propto e^{i\pi r}r^{-1/K}\label{doublet correlation}\\
      &\ev*{S^z_rS^z_0}\propto e^{i\pi r} \ev{\cos\sqrt{2}\phi(r)\cdot\cos\sqrt{2}\phi(0)}\notag\\
            &\quad\quad\propto e^{i\pi r}r^{-K}\label{Neel correlation}\\
      &\ev*{(S^+_rS^-_{r+1}+S^-_rS^+_{r+1})(S^+_0S^-_{1}+S^-_0S^+_{1})}\notag\\
            &\quad\quad\propto e^{i\pi r} \ev{\sin\sqrt{2}\phi(r)\cdot\sin\sqrt{2}\phi(0)}\notag\\
            &\quad\quad\propto e^{i\pi r}r^{-K}\label{Dimer correlation}
\end{align}
For simplicity, in Eq.~(\ref{Neel correlation}), the non-oscillating term $\ev*{\partial_x \phi(r)\partial_x \phi(0)} \propto 1/r^2$ has been omitted.
Here, the fields 
\begin{align}
      \cos \sqrt{2} \phi &= \frac{1}{2} (\mathcal{O}_{0,1}+\mathcal{O}_{0,-1})
      =: \mathcal{O}_{\text{N\'{e}el}}  \label{Neel exitation}\\
      \sin \sqrt{2} \phi &= \frac{1}{2} (\mathcal{O}_{0,1}-\mathcal{O}_{0,-1})
      =: \mathcal{O}_{\rm dimer} \label{Dimer exitation}
\end{align}
are called the N\'{e}el excitation and the dimer excitation, respectively. 
The scaling dimensions of the N\'{e}el and dimer excitations are given as $x_{\text{N\'{e}el}}=x_{\text{dimer}}=K/2$ on the GFLs. 
Additionally, the excitation $e^{i\sqrt{2}\theta}=\mathcal{O}_{1,0}$ in Eq.~(\ref{doublet correlation}) has $m=1$ and $x_{1,0}=1/2K$. 
The symmetries of ground state (GS) and N\'{e}el and dimer excitation excitation were analyzed~\cite{Nomura:1994aa} and the results are summarized as follows
\begin{itemize}
\item For $L=4n$:
\begin{align}
      (m,k,P)=
      \begin{cases}
            (0,  0, 1) 	& 	(\text{GS})\\
            (0,\pi,-1) 	& 	(\text{N\'{e}el})\\
            (0,\pi, 1) 	& 	(\text{dimer})
      \end{cases}
\end{align}
\item For $L=4n+2$:
\begin{align}
      (m,k,P)=
      \begin{cases}
            (0,\pi,-1) 	& 	(\text{GS})\\
            (0,  0, 1) 	& 	(\text{N\'{e}el})\\
            (0,  0,-1) 	& 	(\text{dimer})
      \end{cases}
\end{align}
\end{itemize}

\subsection{Kadanoff-Brown's disscussion}\label{Kadanoff-Brown discussion}

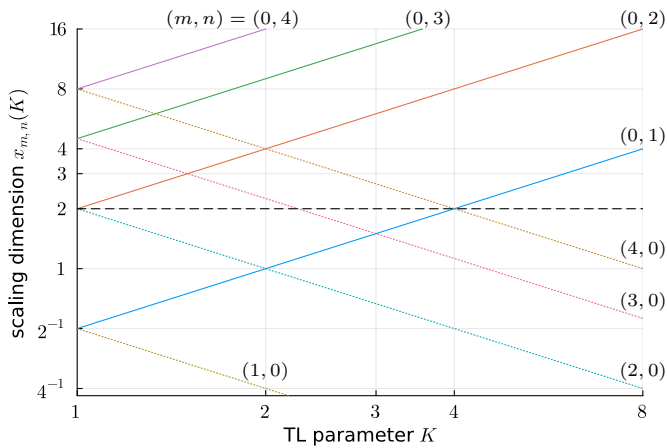
\begin{figure}[tbp]
      \centering
      \begin{tikzpicture}
            \node[anchor=south west,inner sep=0] (image) at (0,0) {
            \includegraphics[width=\linewidth]{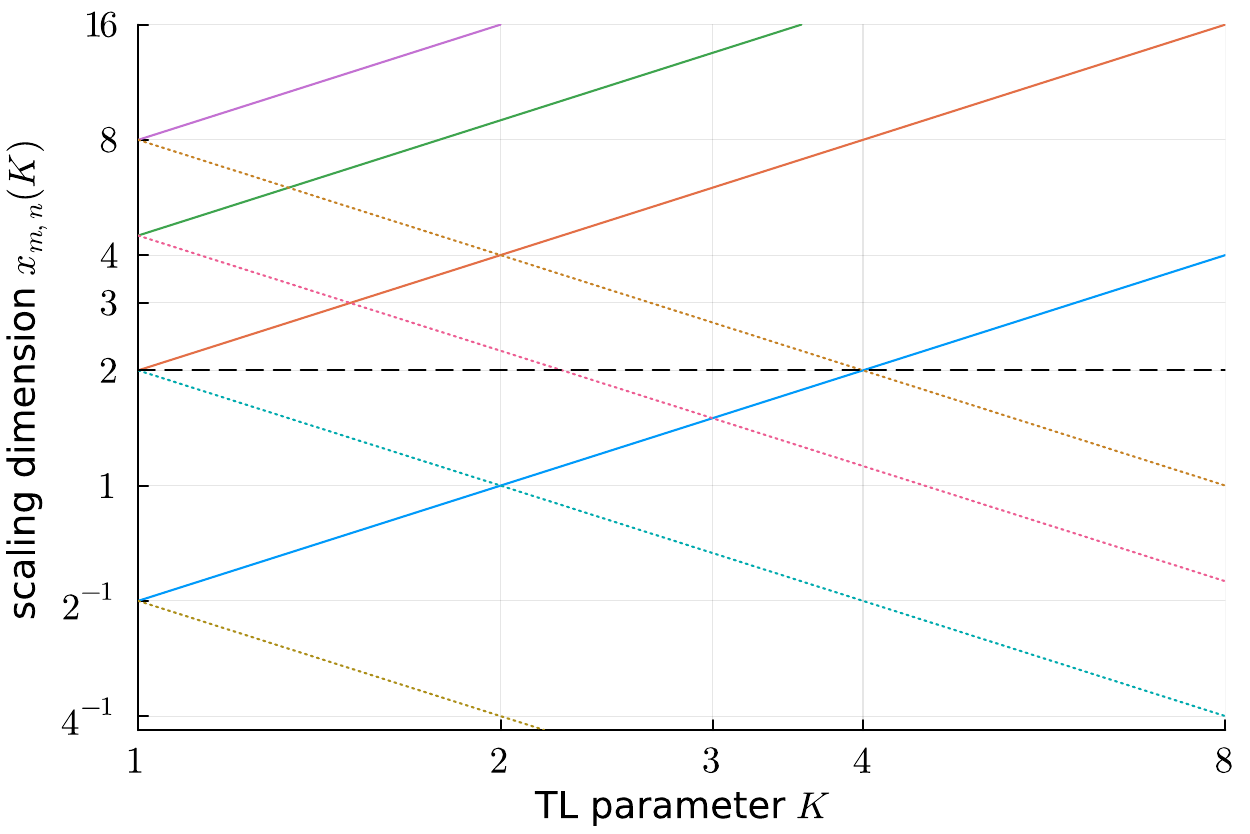}
            };
            \begin{scope}[x={(image.south east)},y={(image.north west)}]
                  \node[anchor=south] at (0.98,0.68) {\scriptsize{$(0,1)$}};
                  \node[anchor=south] at (0.98,0.95) {\scriptsize{$(0,2)$}};
                  \node[anchor=south] at (0.65,0.95) {\scriptsize{$(0,3)$}};
                  \node[anchor=south] at (0.35,0.95) {\scriptsize{$(m,n)=(0,4)$}};
                  \node[anchor=south] at (0.98,0.42) {\scriptsize{$(4,0)$}};
                  \node[anchor=south] at (0.98,0.30) {\scriptsize{$(3,0)$}};
                  \node[anchor=south] at (0.98,0.14) {\scriptsize{$(2,0)$}};
                  \node[anchor=south] at (0.40,0.14) {\scriptsize{$(1,0)$}};
            \end{scope}
      \end{tikzpicture}
      \caption{(Color online)
            The scaling dimensions $x_{m,n}$ of the vertex operators $\mathcal{O}_{m,n}$ in the Gaussian model with $K$.
            The $x_{m,n}$ are plotted as a function of $K$ on log-log scale. 
            When the lines cross, the scaling dimensions $x_{m,n}$ are equal, which corresponds to the level crossing of the vertex operators. 
            The black dashed line represents the line $x_{m,n}=2$, which is the marginal.
      }
      \label{fig: TL parameter}
\end{figure}

Kadanoff and Brown~\cite{KADANOFF1979318} studied the level crossing of the vertex operators in the Gaussian model with $K$. 
For example, $x_{0,1}=x_{2,0}$ for $K=2$, $x_{0,1}=x_{3,0}$ for $K=3$ and $x_{0,1}=x_{4,0}$ for $K=4$, corresponding respectively to the special points $(\Delta,\alpha)=(0,0),(-1/2,0),(-1/\sqrt{2},0)$ in our notation. 
Other cases are shown in Fig.~\ref{fig: TL parameter}. 
This argument is consistent with the level crossing of $sl_2$ loop algebra symmetry in the Hamiltonian~(\ref{eq: hamiltonian}). 
Although not directly related to our content except for $K=4$, they showed that the Gaussian model with $K$ may describe multicritical points, when the crossing of the scaling dimensions is marginal $(x=2)$.

\subsection{Level spectroscopy}\label{sec: level spectroscopy}
When $y_{\phi}\neq0$, correction terms due to the perturbation of the $\cos$ term in the quantum sine-Gordon model~(\ref{sine-Gordon}) are added to the scaling dimensions of the N\'{e}el and dimer excitations~\cite{Nomura:1994aa}
\begin{align}
      x_{\text{N\'{e}el}}=\frac{K}{2}+\frac{y_{\phi}}{2},\quad 
      x_{\text{dimer}}=\frac{K}{2}-\frac{y_{\phi}}{2}.
\end{align}
Then, we get
\begin{align}
      E_{\text{N\'{e}el}}-E_{\text{dimer}}\propto x_{\text{N\'{e}el}}-x_{\text{dimer}}=y_{\phi}.
\end{align}
Thus, the GFLs can be determined by the level crossing of the N\'{e}el and dimer excitations~\cite{Nomura:1994aa,NomuraKitazawa1996}. 
\par
The FM transition line is calculated by the energy gap between the FM state ($m=\pm L/2$) and the ground state in the XY phase becomes zero on this line.

\section{Numerical results}\label{sec: GFL}
\subsection{Gaussian fixed lines}\label{sec: numerical results}

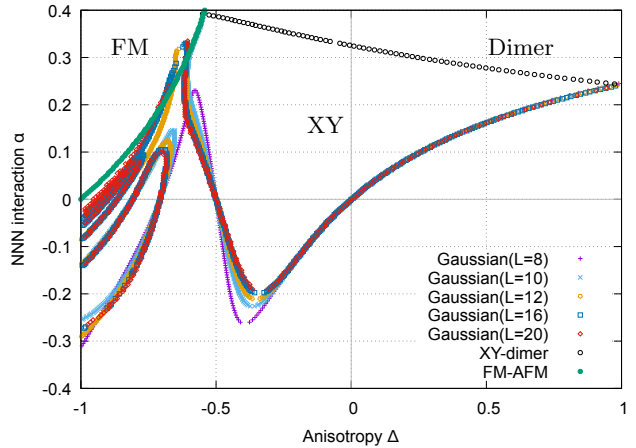
\begin{figure}[tbp]
      \centering
      \begin{tikzpicture}
            \node[anchor=south west,inner sep=0] (duck) at (0,0) {
            \includegraphics[width=\linewidth]{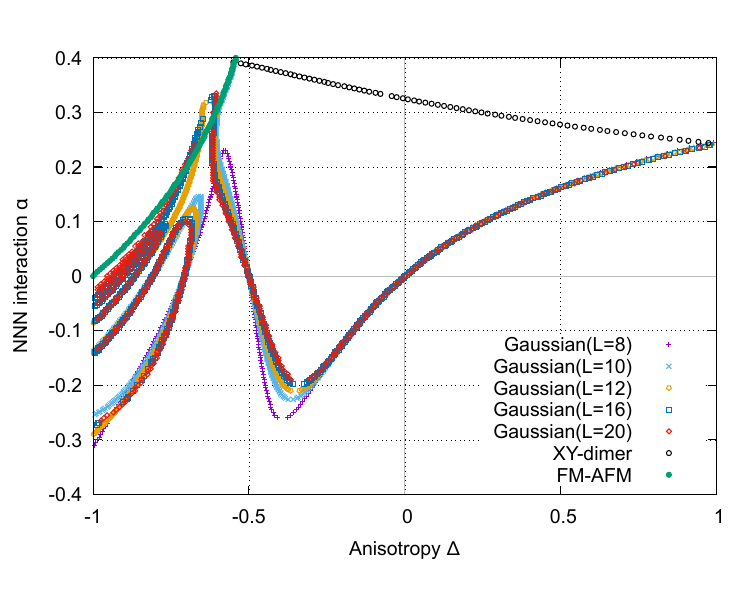}
            };
            \begin{scope}[x={(duck.south east)},y={(duck.north west)}]
                  \node[anchor=south] at (0.2,0.8) {FM};
                  \node[anchor=south] at (0.5,0.65) {XY};
                  \node[anchor=south] at (0.8,0.8) {Dimer};
            \end{scope}
      \end{tikzpicture}
      \caption{(Color online) Gaussian fixed lines for system sizes $L = 8$ to $L=20$.}
      \label{fig:GFLs_L8-20}
\end{figure}

We numerically study the GFLs of the Hamiltonian~(\ref{eq: hamiltonian}) for $L=8$ to $20$ using the level spectroscopy and numerical diagonalizations (see Fig.~\ref{fig:GFLs_L8-20}). 
For $L \le 12$, all energy eigenvalues were computed by full diagonalization using the Householder and inverse iteration methods, whereas for $L > 12$, low-energy eigenvalues were obtained via the Lanczos method with symmetry sectors labeled by magnetization $m$, wave number $k$, and parity $P$. 
All calculations were performed using the quantum spin system diagonalization package TITPACK Ver.~2, partially modified for specific purposes.
\par
The part of our results in the region $\Delta\ge0$ is consistent with the previous study \cite{Nomura:1994aa}. 
Our new result is the clarification of the shape of the GFLs in the region $\Delta \le 0$.
As a consequence, the GFLs pass through all the special points.
In particular, the special points $(\Delta, \alpha) = (0,0)$ and $(-1/2, 0)$, corresponding to $N = 2$ and $N = 3$ in the $sl_2$ loop algebra, respectively, are continuously connected by the GFLs for each system size $L$ we studied. 
Furthermore, Fig.~\ref{fig:GFLs_L8-20} suggests that at $\alpha=0$, the GFLs pass through so as to continuously connect the values of $\Delta$ corresponding to the $K=2n$ and $K=2n+1$ for $n=1,2,\cdots$.

\subsection{Finite size scaling}
\begin{figure}[tbp]
      \centering
      \includegraphics[width=\linewidth]{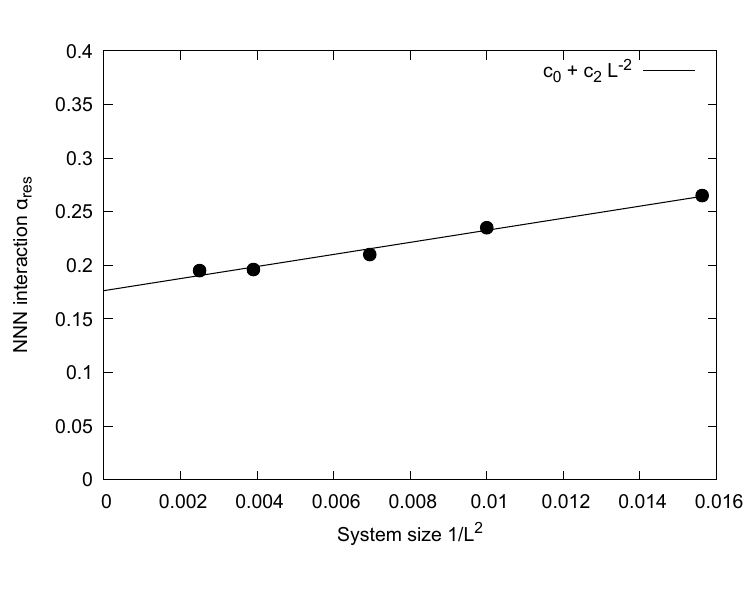}
      \caption{The finite size scaling of $\alpha_{\rm{res}}(L)$.
            It shows the plot of $\alpha_{\rm{res}}(L)$ as a function of $L^{-2}$. 
            The line is the fitting line of Eq.~(\ref{eq: finite size scaling}) and we calculate the extrapolated value of $\alpha_{\rm{res}}(L)$ in the thermodynamic limit. 
            We get the value of $\alpha_{\rm{res}}(\infty)$ is $0.176(3)$. 
      }
      \label{fig:finite_size_scaling}
\end{figure}

In all the system sizes we calculated, the GFL in the region $-1/2\le\Delta\le0$ exhibited similar behavior. 
To discuss whether the meandering of the GFL caused by the $sl_2$ loop algebra also occurs in the thermodynamic limit ($L=\infty$), we focus on the absolute value of the minimum $\alpha$ among the points on the concave fixed line (the ``depth of the valley'') in $-1/2\le\Delta\le0$ (see Fig.~\ref{fig:GFLs_L8-20})
\begin{equation}
      \alpha_{\rm res}= |\min(\alpha)|.
\end{equation}
This value is treated as a system size-dependent quantity $\alpha_{\rm res}(L)$, and finite size scaling was performed to investigate its value in the infinite system (see Fig.~\ref{fig:finite_size_scaling}).
\par

As a result of the calculations, $\alpha_{\rm res}$ was found to decay as
\begin{equation}\label{eq: finite size scaling}
      \alpha_{\rm res}(L) \simeq c_0 + c_2 \times L^{-2},
\end{equation}
and in the thermodynamic limit, it converges to a finite value $\alpha_{\rm res}(\infty) = c_0 = 0.176(3)$ and $c_2=5.6(4)$. 
The reason why the finite size scaling described in Eq.~(\ref{eq: finite size scaling}) is attributed to the descendant field $L_{-2}\bar{L}_{-2}\hat{\bm{1}}$ of the identity operator $\hat{\bm{1}}$. 
This descendant field have scaling dimension $x=4$ and appears as a correction term in the finite-size Hamiltonian~\cite{Cardy:1986aa}. 
This result indicates that the GFLs in the region $-1/2\le\Delta\le0$ are not straight lines, but meandered lines, and the finite size effect of the GFLs shape disappears at the special points caused by the $sl_2$ loop algebra. 

\subsection{TL parameter}\label{sec: TL parameter}

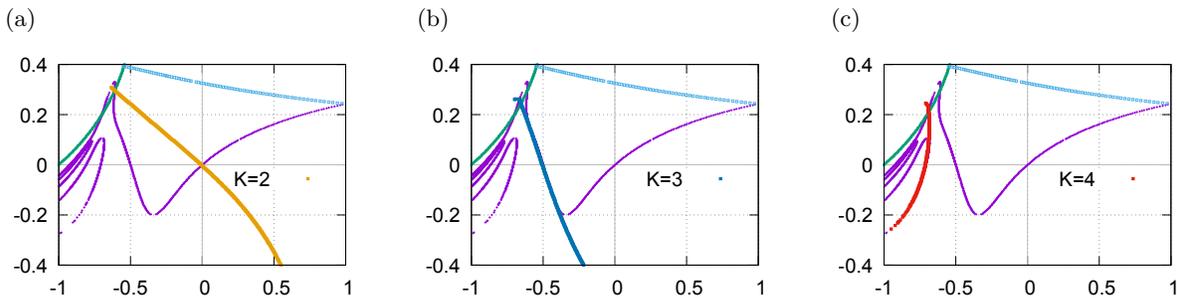
\begin{figure*}[htbp]
      \centering
      \begin{minipage}[t]{0.3\linewidth}
            \begin{tikzpicture}
                  \node[anchor=south west,inner sep=0] (duck) at (0,0) {
                  \includegraphics[width=\linewidth]{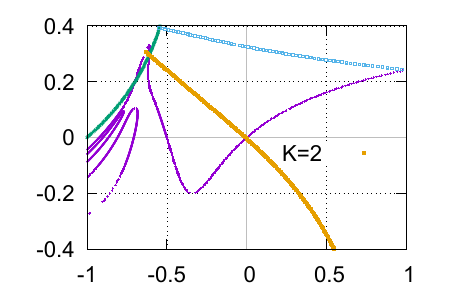}
                  };
                  \begin{scope}[x={(duck.south east)},y={(duck.north west)}]
                        \node[anchor=south] at (0.1,1) {(a)};
                  \end{scope}
            \end{tikzpicture}
      \end{minipage}
      \begin{minipage}[t]{0.3\linewidth}
            \begin{tikzpicture}
                  \node[anchor=south west,inner sep=0] (duck) at (0,0) {
                  \includegraphics[width=\linewidth]{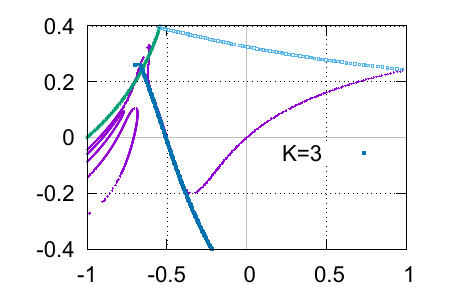}
                  };
                  \begin{scope}[x={(duck.south east)},y={(duck.north west)}]
                        \node[anchor=south] at (0.1,1) {(b)};
                  \end{scope}
            \end{tikzpicture}
      \end{minipage}
      \begin{minipage}[t]{0.3\linewidth}
            \begin{tikzpicture}
                  \node[anchor=south west,inner sep=0] (duck) at (0,0) {
                  \includegraphics[width=\linewidth]{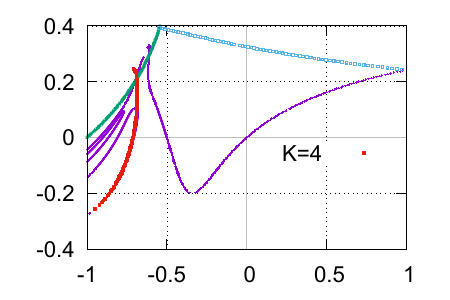}
                  };
                  \begin{scope}[x={(duck.south east)},y={(duck.north west)}]
                        \node[anchor=south] at (0.1,1) {(c)};
                  \end{scope}
            \end{tikzpicture}
      \end{minipage}
      \caption{(Color online) 
            (a), (b) and (c) show the $K=2,3,4$ lines for $L=16$, respectively.
            The XY-dimer transition line is equivalent to the $K=1$ line.
            The $K=2,3,4$ line pass through the special point $(\Delta,\alpha)=(0,0),(-0.5,0),(-1/\sqrt{2},0)$.
      }
      \label{fig:K=2}
\end{figure*}

Hirata and Nomura~\cite{Hirata:2000aa} obtained partial phase diagram of the Hamiltonian (\ref{eq: hamiltonian}) for $\Delta\le0,\alpha\ge0$, and they got the $K=1, 2, 3, 4$ line. 
These lines means the TL parameter $K$ of the quantum sine-Gordon model equals to $1, 2, 3, 4$ on this line. 
The $K=1$ line is equivalent to the XY-dimer transition line, and the transition point in $\Delta=1$ is at $(\Delta,\alpha)=(1,0.241)$ \cite{Okamoto1992}. 
Then this line was calculated by the level spectroscopy method ($E_{\rm dimer} = E_{\pm1}$) to eliminate the logarithmic corrections~\cite{Okamoto1992}. 
The $K=2, 3, 4$ lines were determined by the ratio
\begin{align}
      (x_{\text{N\'{e}el}}+x_{\text{dimer}})/2x_{1,0}=K^2.
\end{align} 
Note that for $K\ge2$, the finite-size correction from the irrelevant term is $L^{2-x}\,(x\ge4)$, thus we ignore it. 
We show the $K=2, 3, 4$ line and the GFLs for $L=16$ in Fig.~\ref{fig:K=2}. 
Fig.~\ref{fig:K=2} (b) and (c) show that the $K=3$ and $K=4$ lines partially overlap with the GFLs. 
This overlap is observed even for $L=20$, suggesting that finite-size effects are small. 
Moreover, the TL parameter $K$ is finite at the point $(\Delta,\alpha)=(-0.61,0.296)$, which is the intersection of the FM transition line, the GFL, and the $K=2$ line (Fig.~\ref{fig:K=2}(a)). 
This result indicates that the TL parameter $K$ changes as $K=1\to2\to3\to2$ along the GFL. 

\section{Discussion}\label{sec: discussion}

\subsection{Metamagnetic region}\label{sec: metamagnetic phase}
\begin{figure}[tbp]
      \centering
      \includegraphics[width=\linewidth]{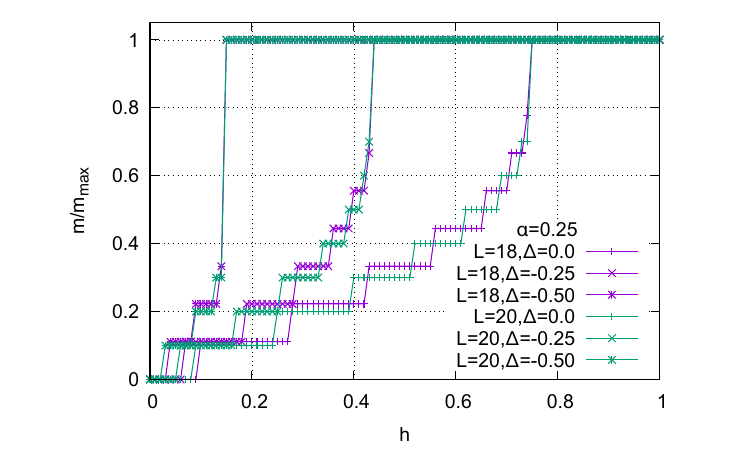}
      \caption{(Color online) 
            The magnetization curve of the Hamiltonian (\ref{eq: hamiltonian}) with $L=18,20,\, \alpha=0.25$. 
            For $\Delta=-0.25,-0.5$ cases, the magnetization curve has a discontinuity at the critical longitudinal magnetic field $h_c(-0.5,0.25)=0.15,h_c(-0.25,0.25)=0.43$ where the magnetization curve jumps to $m=m_{\rm max}:=L/2$.
      }
      \label{fig:mag_curve}
\end{figure}

Gerhardt {\it et al.} and Hirata~\cite{Gerhardt:1998aa,hirata1999} have shown that the Hamiltonian (\ref{eq: hamiltonian}) exhibits a metamagnetic region, which is a part of the XY phase and the dimer phase. 
This region is characterized by the magnetization jump, which has a discontinuity at the critical longitudinal magnetic field $h=h_c(\Delta,\alpha)$ where the magnetization curve jumps from the critical magnetization $m=m_c(\Delta,\alpha)$ to $m=m_{\rm max}:=L/2$. 
So the metamagnetic region is characterized by 
\begin{align}\label{eq: metamagnetic}
      \frac{\partial^2 E_0(m,\Delta,\alpha)}{\partial m^2}
      \begin{cases}
       > 0\quad (0<m<m_c)\\
       = 0\quad (m=m_c)
      \end{cases}
\end{align}
where $E_0(m,\Delta,\alpha)$ is the lowest eigen energy in the magnetization $m$ sector. 
The boundary of metamagnetic region is given by
\begin{align}\label{eq: phase boundary of metamagnetic}
      \left.\frac{\partial^2 E_0(m,\Delta,\alpha)}{\partial m^2}\right|_{m=m_{\rm max}}=0.
\end{align} 
In the numerical calculation, Eq.~(\ref{eq: metamagnetic}) gets negative value for $m_c<m<m_{\rm max}$.
The magnetization curve of the Hamiltonian (\ref{eq: hamiltonian}) is shown in Fig.~\ref{fig:mag_curve}. 
\par

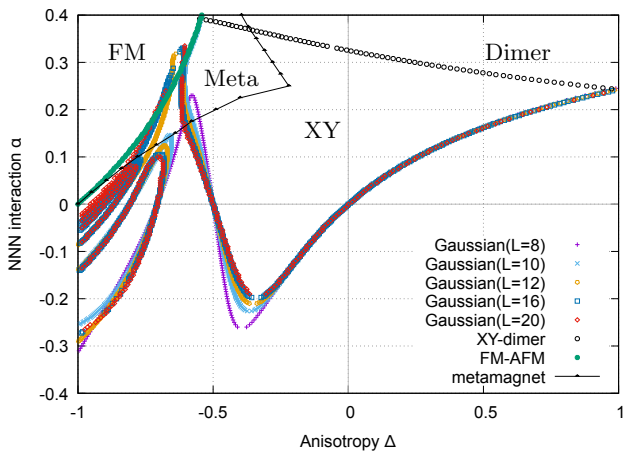
\begin{figure}[tbp]
      \begin{tikzpicture}
            \node[anchor=south west,inner sep=0] (duck) at (0,0) {
            \includegraphics[width=\linewidth]{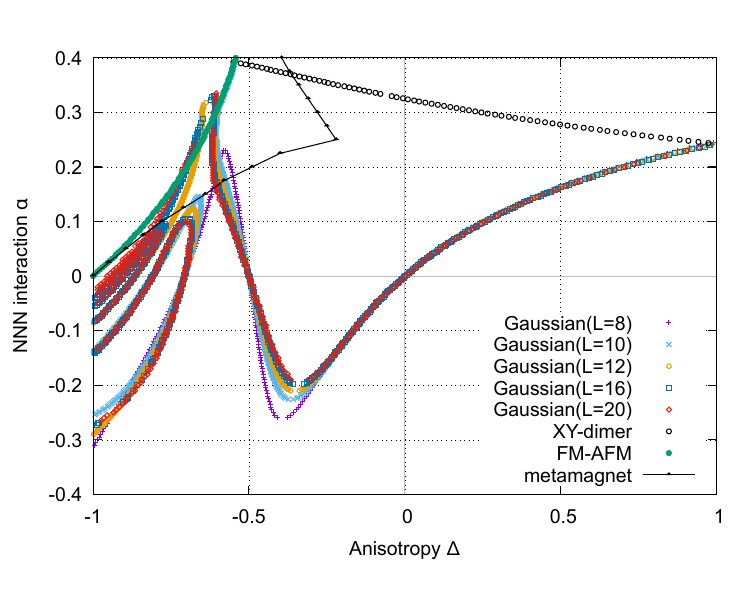}
            };
            \begin{scope}[x={(duck.south east)},y={(duck.north west)}]
                  \node[anchor=south] at (0.2,0.8) {FM};
                  \node[anchor=south] at (0.5,0.65) {XY};
                  \node[anchor=south] at (0.8,0.8) {Dimer};
                  \node[anchor=south] at (0.36,0.75) {Meta};
            \end{scope}
      \end{tikzpicture}
      \caption{(Color online) The GFLs and the metamagnetic region.}
      \label{fig:metamagnet}
\end{figure}

Fig.~\ref{fig:metamagnet} shows the result of calculating the metamagnetic region using Hirata's method \cite{hirata1999}, overlaid with the GFLs, the FM transition line and the XY-dimer transition line. 
It indicates that the behavior of the GFLs differs inside and outside of the metamagnetic region. 
The GFLs, except for passing through special points $N=2,3$ case and $N=L/2$ case, are found to avoid entering the metamagnetic region for each system size.

\subsection{CFT in the metamagnetic region}\label{sec: CFT in the metamagnetic phase}
In this section, we discuss the metamagnetic region and the TL parameter.
\par
When the physical system is written by the $c=1$ CFT, the scaling dimension of the operator $\mathcal{O}_{m,n}$ is given by Eq.~(\ref{eq: scaling dimension with gaussian}) and the excitation gap is given by Eq.~(\ref{eq: energy gap}).
The second derivative Eq.~(\ref{eq: metamagnetic}) is given by
\begin{align}
      \frac{\partial E_{m,n}}{\partial m}=\frac{2\pi vm}{LK}\ge0,\\
      \frac{\partial^2 E_{m,n}}{\partial m^2}=\frac{2\pi v}{LK}\ge0.
\end{align}
This indicates that $E_0(m,\Delta,\alpha)$ is a monotonically increasing and concave function for $m$ in the XY phase, except for $v=0$ or $K=\infty$. 
\par
In the metamagnetic region, the TLL description is expected to remain valid at energy scales where Eq.~(\ref{eq: metamagnetic}) is positive, while it breaks down when the second derivative becomes zero or negative. 
Therefore, the TLL description is valid in the energy range $E_0$ to $E_0+h_c m_c$, where $E_0$ is the ground state energy. 
\par
When $\alpha=0$, Nakamura and Nomura~\cite{Nakamura:1997aa} showed that the TL parameter $K$ increases monotonically as $\Delta$ approaches the FM transition point, where it diverges. 
Comparing this fact with our results, we find that the behavior of the TL parameter on the FM trasition point is different for $\alpha=0$ and $\alpha\ne0$. 

\section{Conclusion} \label{sec: conclusion}
We have studied the GFLs of the Hamiltonian~(\ref{eq: hamiltonian}) in the critical phase using level spectroscopy and numerical diagonalization. 
On the GFLs, the low-energy effective field theory is described by the Gaussian model~(\ref{eq: hamiltonian gaussian}).
The Gaussian model is known to the CFT with the central charge $c=1$. This model do not have the umklapp scattering term compared to the quantum sine-Gordon model, which is the low-energy effective field theory of the Hamiltonian~(\ref{eq: hamiltonian}). 
In this paper, we examine the relation between the GFLs and the special points, which are defined as the points where the $sl_2$ loop algebra symmetry governs in low-energy physics. 
\par
First, we have numerically shown that the GFLs pass through the special points, and some of the GFLs are connected. 
It was already known that there exist infinitely many GFLs in the $S=1/2$ XXZ chain with FM-NN interactions and AFM-NNN interactions~\cite{Furukawa:2010aa}, but the possibility of connections of the GFLs was not considered. 
Besides, our results are consistent with the anomalous behavior of the non-linear Drude weight in the $S=1/2$ XXZ chain~\cite{Tanikawa:2021ab}, which is thought to be caused by the umklapp term vanishing at the special points.
\par
Next, we calculated the finite-size scaling for the shape of the GFLs in the region of $-1/2\le\Delta\le0$.
This calculation indicates that the GFLs are not straight in the thermodynamic limit, but exhibit meandering behavior due to the $sl_2$ loop algebra symmetry.
Then we found that the finite-size effects vanish at the special points. 
\par
In addition, we calculated the TL parameter $K$ of the quantum sine-Gordon model. 
We found that it appears that the $K=3$ and $K=4$ lines partially overlap with the GFLs. 
The physical meaning of this result is future work.
\par
Finally, we discussed the shape of the GFLs for $-1\le\Delta<-0$ reflects the metamagnetic region.
We found that the GFLs do not enter the metamagnetic region except for passing through the special points $N=2,3$ case and $N=L/2$ case.
This suggests that the TLL description is invarid in the metamagnetic region.

% Acknowledgments are optional
%\begin{acknowledgments}
\acknowledgments
We would like to thank T. Deguchi, H. Katsura, C. Matsui and S. Furukawa for useful discussions. 
K. N. is supported by Japan Society for the Promotion of Science KAKENHI Grant No. 21H05021. D.Y. was supported by JST SPRING, Japan Grant Number JPMJSP2136. 
The numerical calculation in this work was based on the TITPACK version 2, developed by Professor H. Nishimori. 
%\end{acknowledgments}
%\appendix
%\section{}

%\begin{acknowledgments}
%\end{acknowledgments}

\bibliographystyle{apsrev4-2}
\bibliography{reference.bib}

\begin{thebibliography}{31}%
\makeatletter
\providecommand \@ifxundefined [1]{%
 \@ifx{#1\undefined}
}%
\providecommand \@ifnum [1]{%
 \ifnum #1\expandafter \@firstoftwo
 \else \expandafter \@secondoftwo
 \fi
}%
\providecommand \@ifx [1]{%
 \ifx #1\expandafter \@firstoftwo
 \else \expandafter \@secondoftwo
 \fi
}%
\providecommand \natexlab [1]{#1}%
\providecommand \enquote  [1]{``#1''}%
\providecommand \bibnamefont  [1]{#1}%
\providecommand \bibfnamefont [1]{#1}%
\providecommand \citenamefont [1]{#1}%
\providecommand \href@noop [0]{\@secondoftwo}%
\providecommand \href [0]{\begingroup \@sanitize@url \@href}%
\providecommand \@href[1]{\@@startlink{#1}\@@href}%
\providecommand \@@href[1]{\endgroup#1\@@endlink}%
\providecommand \@sanitize@url [0]{\catcode `\\12\catcode `\$12\catcode
  `\&12\catcode `\#12\catcode `\^12\catcode `\_12\catcode `\%12\relax}%
\providecommand \@@startlink[1]{}%
\providecommand \@@endlink[0]{}%
\providecommand \url  [0]{\begingroup\@sanitize@url \@url }%
\providecommand \@url [1]{\endgroup\@href {#1}{\urlprefix }}%
\providecommand \urlprefix  [0]{URL }%
\providecommand \Eprint [0]{\href }%
\providecommand \doibase [0]{https://doi.org/}%
\providecommand \selectlanguage [0]{\@gobble}%
\providecommand \bibinfo  [0]{\@secondoftwo}%
\providecommand \bibfield  [0]{\@secondoftwo}%
\providecommand \translation [1]{[#1]}%
\providecommand \BibitemOpen [0]{}%
\providecommand \bibitemStop [0]{}%
\providecommand \bibitemNoStop [0]{.\EOS\space}%
\providecommand \EOS [0]{\spacefactor3000\relax}%
\providecommand \BibitemShut  [1]{\csname bibitem#1\endcsname}%
\let\auto@bib@innerbib\@empty
%</preamble>
\bibitem [{\citenamefont {Ashcroft}\ and\ \citenamefont
  {Mermin}(1976)}]{Ashcroft-Mermin:1976aa}%
  \BibitemOpen
  \bibfield  {author} {\bibinfo {author} {\bibfnamefont {N.}~\bibnamefont
  {Ashcroft}}\ and\ \bibinfo {author} {\bibfnamefont {N.}~\bibnamefont
  {Mermin}},\ }\href {https://books.google.com/books?id=1C9HAQAAIAAJ} {\emph
  {\bibinfo {title} {Solid State Physics}}},\ HRW international editions\
  (\bibinfo  {publisher} {Holt, Rinehart and Winston},\ \bibinfo {year}
  {1976})\BibitemShut {NoStop}%
\bibitem [{\citenamefont {Tanikawa}\ \emph {et~al.}(2021)\citenamefont
  {Tanikawa}, \citenamefont {Takasan},\ and\ \citenamefont
  {Katsura}}]{Tanikawa:2021aa}%
  \BibitemOpen
  \bibfield  {author} {\bibinfo {author} {\bibfnamefont {Y.}~\bibnamefont
  {Tanikawa}}, \bibinfo {author} {\bibfnamefont {K.}~\bibnamefont {Takasan}},\
  and\ \bibinfo {author} {\bibfnamefont {H.}~\bibnamefont {Katsura}},\ }\href
  {https://doi.org/10.1103/PhysRevB.103.L201120} {\bibfield  {journal}
  {\bibinfo  {journal} {Physical Review B}\ }\textbf {\bibinfo {volume}
  {103}},\ \bibinfo {pages} {L201120} (\bibinfo {year} {2021})}\BibitemShut
  {NoStop}%
\bibitem [{\citenamefont {Tanikawa}\ and\ \citenamefont
  {Katsura}(2021)}]{Tanikawa:2021ab}%
  \BibitemOpen
  \bibfield  {author} {\bibinfo {author} {\bibfnamefont {Y.}~\bibnamefont
  {Tanikawa}}\ and\ \bibinfo {author} {\bibfnamefont {H.}~\bibnamefont
  {Katsura}},\ }\href {https://doi.org/10.1103/PhysRevB.104.205116} {\bibfield
  {journal} {\bibinfo  {journal} {Physical Review B}\ }\textbf {\bibinfo
  {volume} {104}},\ \bibinfo {pages} {205116} (\bibinfo {year}
  {2021})}\BibitemShut {NoStop}%
\bibitem [{\citenamefont {Giamarchi}(2003)}]{Giamarchi:2003aa}%
  \BibitemOpen
  \bibfield  {author} {\bibinfo {author} {\bibfnamefont {T.}~\bibnamefont
  {Giamarchi}},\ }in\ \href
  {https://doi.org/10.1093/acprof:oso/9780198525004.003.0002} {\emph {\bibinfo
  {booktitle} {Quantum Physics in One Dimension}}}\ (\bibinfo  {publisher}
  {Oxford University Press},\ \bibinfo {year} {2003})\BibitemShut {NoStop}%
\bibitem [{\citenamefont {Bethe}(1931)}]{Bethe:1931aa}%
  \BibitemOpen
  \bibfield  {author} {\bibinfo {author} {\bibfnamefont {H.}~\bibnamefont
  {Bethe}},\ }\href {https://doi.org/10.1007/BF01341708} {\bibfield  {journal}
  {\bibinfo  {journal} {Zeitschrift f{\"u}r Physik}\ }\textbf {\bibinfo
  {volume} {71}},\ \bibinfo {pages} {205} (\bibinfo {year} {1931})}\BibitemShut
  {NoStop}%
\bibitem [{\citenamefont {Des~Cloizeaux}\ and\ \citenamefont
  {Gaudin}(1966)}]{Des-Cloizeaux:1966aa}%
  \BibitemOpen
  \bibfield  {author} {\bibinfo {author} {\bibfnamefont {J.}~\bibnamefont
  {Des~Cloizeaux}}\ and\ \bibinfo {author} {\bibfnamefont {M.}~\bibnamefont
  {Gaudin}},\ }\href {https://doi.org/10.1063/1.1705048} {\bibfield  {journal}
  {\bibinfo  {journal} {Journal of Mathematical Physics}\ }\textbf {\bibinfo
  {volume} {7}},\ \bibinfo {pages} {1384} (\bibinfo {year} {1966})}\BibitemShut
  {NoStop}%
\bibitem [{\citenamefont {Lieb}\ \emph {et~al.}(1961)\citenamefont {Lieb},
  \citenamefont {Schultz},\ and\ \citenamefont {Mattis}}]{Lieb:1961aa}%
  \BibitemOpen
  \bibfield  {author} {\bibinfo {author} {\bibfnamefont {E.}~\bibnamefont
  {Lieb}}, \bibinfo {author} {\bibfnamefont {T.}~\bibnamefont {Schultz}},\ and\
  \bibinfo {author} {\bibfnamefont {D.}~\bibnamefont {Mattis}},\ }\href
  {https://doi.org/https://doi.org/10.1016/0003-4916(61)90115-4} {\bibfield
  {journal} {\bibinfo  {journal} {Annals of Physics}\ }\textbf {\bibinfo
  {volume} {16}},\ \bibinfo {pages} {407} (\bibinfo {year} {1961})}\BibitemShut
  {NoStop}%
\bibitem [{\citenamefont {Tonegawa}\ and\ \citenamefont
  {Harada}(1987)}]{Tonegawa1987:aa}%
  \BibitemOpen
  \bibfield  {author} {\bibinfo {author} {\bibfnamefont {T.}~\bibnamefont
  {Tonegawa}}\ and\ \bibinfo {author} {\bibfnamefont {I.}~\bibnamefont
  {Harada}},\ }\href {https://doi.org/10.1143/JPSJ.56.2153} {\bibfield
  {journal} {\bibinfo  {journal} {Journal of the Physical Society of Japan}\
  }\textbf {\bibinfo {volume} {56}},\ \bibinfo {pages} {2153} (\bibinfo {year}
  {1987})},\ \Eprint
  {https://arxiv.org/abs/https://doi.org/10.1143/JPSJ.56.2153}
  {https://doi.org/10.1143/JPSJ.56.2153} \BibitemShut {NoStop}%
\bibitem [{\citenamefont {Okamoto}\ and\ \citenamefont
  {Nomura}(1992)}]{Okamoto1992}%
  \BibitemOpen
  \bibfield  {author} {\bibinfo {author} {\bibfnamefont {K.}~\bibnamefont
  {Okamoto}}\ and\ \bibinfo {author} {\bibfnamefont {K.}~\bibnamefont
  {Nomura}},\ }\href
  {https://doi.org/https://doi.org/10.1016/0375-9601(92)90823-5} {\bibfield
  {journal} {\bibinfo  {journal} {Physics Letters A}\ }\textbf {\bibinfo
  {volume} {169}},\ \bibinfo {pages} {433} (\bibinfo {year}
  {1992})}\BibitemShut {NoStop}%
\bibitem [{\citenamefont {Farnell}\ and\ \citenamefont
  {Parkinson}(1994)}]{Farnell1994:aa}%
  \BibitemOpen
  \bibfield  {author} {\bibinfo {author} {\bibfnamefont {D.~J.~J.}\
  \bibnamefont {Farnell}}\ and\ \bibinfo {author} {\bibfnamefont {J.~B.}\
  \bibnamefont {Parkinson}},\ }\href
  {https://doi.org/10.1088/0953-8984/6/28/024} {\bibfield  {journal} {\bibinfo
  {journal} {Journal of Physics: Condensed Matter}\ }\textbf {\bibinfo {volume}
  {6}},\ \bibinfo {pages} {5521} (\bibinfo {year} {1994})}\BibitemShut
  {NoStop}%
\bibitem [{\citenamefont {Majumdar}\ and\ \citenamefont
  {Ghosh}(1969{\natexlab{a}})}]{Majumdar:1969aa}%
  \BibitemOpen
  \bibfield  {author} {\bibinfo {author} {\bibfnamefont {C.~K.}\ \bibnamefont
  {Majumdar}}\ and\ \bibinfo {author} {\bibfnamefont {D.~K.}\ \bibnamefont
  {Ghosh}},\ }\href {https://doi.org/10.1063/1.1664978} {\bibfield  {journal}
  {\bibinfo  {journal} {Journal of Mathematical Physics}\ }\textbf {\bibinfo
  {volume} {10}},\ \bibinfo {pages} {1388} (\bibinfo {year}
  {1969}{\natexlab{a}})}\BibitemShut {NoStop}%
\bibitem [{\citenamefont {Majumdar}\ and\ \citenamefont
  {Ghosh}(1969{\natexlab{b}})}]{Majumdar:1969ab}%
  \BibitemOpen
  \bibfield  {author} {\bibinfo {author} {\bibfnamefont {C.~K.}\ \bibnamefont
  {Majumdar}}\ and\ \bibinfo {author} {\bibfnamefont {D.~K.}\ \bibnamefont
  {Ghosh}},\ }\href {https://doi.org/10.1063/1.1664979} {\bibfield  {journal}
  {\bibinfo  {journal} {Journal of Mathematical Physics}\ }\textbf {\bibinfo
  {volume} {10}},\ \bibinfo {pages} {1399} (\bibinfo {year}
  {1969}{\natexlab{b}})}\BibitemShut {NoStop}%
\bibitem [{\citenamefont {{van den Broek}}(1980)}]{VANDENBROEK1980261}%
  \BibitemOpen
  \bibfield  {author} {\bibinfo {author} {\bibfnamefont {P.}~\bibnamefont {{van
  den Broek}}},\ }\href
  {https://doi.org/https://doi.org/10.1016/0375-9601(80)90662-3} {\bibfield
  {journal} {\bibinfo  {journal} {Physics Letters A}\ }\textbf {\bibinfo
  {volume} {77}},\ \bibinfo {pages} {261} (\bibinfo {year} {1980})}\BibitemShut
  {NoStop}%
\bibitem [{\citenamefont {Nomura}\ and\ \citenamefont
  {Okamoto}(1993)}]{Nomura:1993aa}%
  \BibitemOpen
  \bibfield  {author} {\bibinfo {author} {\bibfnamefont {K.}~\bibnamefont
  {Nomura}}\ and\ \bibinfo {author} {\bibfnamefont {K.}~\bibnamefont
  {Okamoto}},\ }\href {https://doi.org/10.1143/JPSJ.62.1123} {\bibfield
  {journal} {\bibinfo  {journal} {Journal of the Physical Society of Japan}\
  }\textbf {\bibinfo {volume} {62}},\ \bibinfo {pages} {1123} (\bibinfo {year}
  {1993})}\BibitemShut {NoStop}%
\bibitem [{\citenamefont {Hirata}\ and\ \citenamefont
  {Nomura}(2000)}]{Hirata:2000aa}%
  \BibitemOpen
  \bibfield  {author} {\bibinfo {author} {\bibfnamefont {S.}~\bibnamefont
  {Hirata}}\ and\ \bibinfo {author} {\bibfnamefont {K.}~\bibnamefont
  {Nomura}},\ }\href {https://doi.org/10.1103/PhysRevB.61.9453} {\bibfield
  {journal} {\bibinfo  {journal} {Physical Review B}\ }\textbf {\bibinfo
  {volume} {61}},\ \bibinfo {pages} {9453} (\bibinfo {year}
  {2000})}\BibitemShut {NoStop}%
\bibitem [{\citenamefont {Furukawa}\ \emph {et~al.}(2010)\citenamefont
  {Furukawa}, \citenamefont {Sato},\ and\ \citenamefont
  {Furusaki}}]{Furukawa:2010aa}%
  \BibitemOpen
  \bibfield  {author} {\bibinfo {author} {\bibfnamefont {S.}~\bibnamefont
  {Furukawa}}, \bibinfo {author} {\bibfnamefont {M.}~\bibnamefont {Sato}},\
  and\ \bibinfo {author} {\bibfnamefont {A.}~\bibnamefont {Furusaki}},\ }\href
  {https://doi.org/10.1103/PhysRevB.81.094430} {\bibfield  {journal} {\bibinfo
  {journal} {Physical Review B - Condensed Matter and Materials Physics}\
  }\textbf {\bibinfo {volume} {81}},\ \bibinfo {pages} {094430} (\bibinfo
  {year} {2010})}\BibitemShut {NoStop}%
\bibitem [{\citenamefont {Nakamura}\ and\ \citenamefont
  {Nomura}(1997)}]{Nakamura:1997aa}%
  \BibitemOpen
  \bibfield  {author} {\bibinfo {author} {\bibfnamefont {M.}~\bibnamefont
  {Nakamura}}\ and\ \bibinfo {author} {\bibfnamefont {K.}~\bibnamefont
  {Nomura}},\ }\href {https://doi.org/10.1103/PhysRevB.56.12840} {\bibfield
  {journal} {\bibinfo  {journal} {Physical Review B}\ }\textbf {\bibinfo
  {volume} {56}},\ \bibinfo {pages} {12840} (\bibinfo {year}
  {1997})}\BibitemShut {NoStop}%
\bibitem [{\citenamefont {Dutton}\ \emph {et~al.}(2012)\citenamefont {Dutton},
  \citenamefont {Kumar}, \citenamefont {Mourigal}, \citenamefont {Soos},
  \citenamefont {Wen}, \citenamefont {Broholm}, \citenamefont {Andersen},
  \citenamefont {Huang}, \citenamefont {Zbiri}, \citenamefont {Toft-Petersen},\
  and\ \citenamefont {Cava}}]{Dutton:2012aa}%
  \BibitemOpen
  \bibfield  {author} {\bibinfo {author} {\bibfnamefont {S.~E.}\ \bibnamefont
  {Dutton}}, \bibinfo {author} {\bibfnamefont {M.}~\bibnamefont {Kumar}},
  \bibinfo {author} {\bibfnamefont {M.}~\bibnamefont {Mourigal}}, \bibinfo
  {author} {\bibfnamefont {Z.~G.}\ \bibnamefont {Soos}}, \bibinfo {author}
  {\bibfnamefont {J.-J.}\ \bibnamefont {Wen}}, \bibinfo {author} {\bibfnamefont
  {C.~L.}\ \bibnamefont {Broholm}}, \bibinfo {author} {\bibfnamefont {N.~H.}\
  \bibnamefont {Andersen}}, \bibinfo {author} {\bibfnamefont {Q.}~\bibnamefont
  {Huang}}, \bibinfo {author} {\bibfnamefont {M.}~\bibnamefont {Zbiri}},
  \bibinfo {author} {\bibfnamefont {R.}~\bibnamefont {Toft-Petersen}},\ and\
  \bibinfo {author} {\bibfnamefont {R.~J.}\ \bibnamefont {Cava}},\ }\href
  {https://doi.org/10.1103/PhysRevLett.108.187206} {\bibfield  {journal}
  {\bibinfo  {journal} {Phys. Rev. Lett.}\ }\textbf {\bibinfo {volume} {108}},\
  \bibinfo {pages} {187206} (\bibinfo {year} {2012})}\BibitemShut {NoStop}%
\bibitem [{\citenamefont {Nomura}\ and\ \citenamefont
  {Okamoto}(1994)}]{Nomura:1994aa}%
  \BibitemOpen
  \bibfield  {author} {\bibinfo {author} {\bibfnamefont {K.}~\bibnamefont
  {Nomura}}\ and\ \bibinfo {author} {\bibfnamefont {K.}~\bibnamefont
  {Okamoto}},\ }\href {https://doi.org/10.1088/0305-4470/27/17/012} {\bibfield
  {journal} {\bibinfo  {journal} {Journal of Physics A: Mathematical and
  General}\ }\textbf {\bibinfo {volume} {27}},\ \bibinfo {pages} {5773}
  (\bibinfo {year} {1994})}\BibitemShut {NoStop}%
\bibitem [{\citenamefont {Deguchi}\ \emph {et~al.}(2001)\citenamefont
  {Deguchi}, \citenamefont {Fabricius},\ and\ \citenamefont
  {McCoy}}]{Deguchi:2001aa}%
  \BibitemOpen
  \bibfield  {author} {\bibinfo {author} {\bibfnamefont {T.}~\bibnamefont
  {Deguchi}}, \bibinfo {author} {\bibfnamefont {K.}~\bibnamefont {Fabricius}},\
  and\ \bibinfo {author} {\bibfnamefont {B.~M.}\ \bibnamefont {McCoy}},\ }\href
  {https://doi.org/10.1023/A:1004894701900} {\bibfield  {journal} {\bibinfo
  {journal} {Journal of Statistical Physics}\ }\textbf {\bibinfo {volume}
  {102}},\ \bibinfo {pages} {701} (\bibinfo {year} {2001})}\BibitemShut
  {NoStop}%
\bibitem [{\citenamefont {Gerhardt}\ \emph {et~al.}(1998)\citenamefont
  {Gerhardt}, \citenamefont {M\"utter},\ and\ \citenamefont
  {Kr\"oger}}]{Gerhardt:1998aa}%
  \BibitemOpen
  \bibfield  {author} {\bibinfo {author} {\bibfnamefont {C.}~\bibnamefont
  {Gerhardt}}, \bibinfo {author} {\bibfnamefont {K.-H.}\ \bibnamefont
  {M\"utter}},\ and\ \bibinfo {author} {\bibfnamefont {H.}~\bibnamefont
  {Kr\"oger}},\ }\href {https://doi.org/10.1103/PhysRevB.57.11504} {\bibfield
  {journal} {\bibinfo  {journal} {Phys. Rev. B}\ }\textbf {\bibinfo {volume}
  {57}},\ \bibinfo {pages} {11504} (\bibinfo {year} {1998})}\BibitemShut
  {NoStop}%
\bibitem [{\citenamefont {Pasquier}\ and\ \citenamefont
  {Saleur}(1990)}]{Pasquier:1990aa}%
  \BibitemOpen
  \bibfield  {author} {\bibinfo {author} {\bibfnamefont {V.}~\bibnamefont
  {Pasquier}}\ and\ \bibinfo {author} {\bibfnamefont {H.}~\bibnamefont
  {Saleur}},\ }\href
  {https://doi.org/https://doi.org/10.1016/0550-3213(90)90122-T} {\bibfield
  {journal} {\bibinfo  {journal} {Nuclear Physics B}\ }\textbf {\bibinfo
  {volume} {330}},\ \bibinfo {pages} {523} (\bibinfo {year}
  {1990})}\BibitemShut {NoStop}%
\bibitem [{\citenamefont {Moudgalya}\ \emph {et~al.}(2020)\citenamefont
  {Moudgalya}, \citenamefont {Regnault},\ and\ \citenamefont
  {Bernevig}}]{Moudgalya:2020aa}%
  \BibitemOpen
  \bibfield  {author} {\bibinfo {author} {\bibfnamefont {S.}~\bibnamefont
  {Moudgalya}}, \bibinfo {author} {\bibfnamefont {N.}~\bibnamefont
  {Regnault}},\ and\ \bibinfo {author} {\bibfnamefont {B.~A.}\ \bibnamefont
  {Bernevig}},\ }\href {https://doi.org/10.1103/PhysRevB.102.085140} {\bibfield
   {journal} {\bibinfo  {journal} {Phys. Rev. B}\ }\textbf {\bibinfo {volume}
  {102}},\ \bibinfo {pages} {085140} (\bibinfo {year} {2020})}\BibitemShut
  {NoStop}%
\bibitem [{\citenamefont {Jordan}\ and\ \citenamefont
  {Wigner}(1928)}]{Jordan:1928}%
  \BibitemOpen
  \bibfield  {author} {\bibinfo {author} {\bibfnamefont {P.}~\bibnamefont
  {Jordan}}\ and\ \bibinfo {author} {\bibfnamefont {E.}~\bibnamefont
  {Wigner}},\ }\href {https://doi.org/10.1007/BF01331938} {\bibfield  {journal}
  {\bibinfo  {journal} {Zeitschrift f{\"u}r Physik}\ }\textbf {\bibinfo
  {volume} {47}},\ \bibinfo {pages} {631} (\bibinfo {year} {1928})}\BibitemShut
  {NoStop}%
\bibitem [{\citenamefont {Haldane}(1982)}]{Haldane:1982aa}%
  \BibitemOpen
  \bibfield  {author} {\bibinfo {author} {\bibfnamefont {F.~D.~M.}\
  \bibnamefont {Haldane}},\ }\href {https://doi.org/10.1103/PhysRevB.25.4925}
  {\bibfield  {journal} {\bibinfo  {journal} {Physical Review B}\ }\textbf
  {\bibinfo {volume} {25}},\ \bibinfo {pages} {4925} (\bibinfo {year}
  {1982})}\BibitemShut {NoStop}%
\bibitem [{\citenamefont {Amit}\ \emph {et~al.}(1980)\citenamefont {Amit},
  \citenamefont {Goldschmidt},\ and\ \citenamefont {Grinstein}}]{Amit:1980aa}%
  \BibitemOpen
  \bibfield  {author} {\bibinfo {author} {\bibfnamefont {D.~J.}\ \bibnamefont
  {Amit}}, \bibinfo {author} {\bibfnamefont {Y.~Y.}\ \bibnamefont
  {Goldschmidt}},\ and\ \bibinfo {author} {\bibfnamefont {S.}~\bibnamefont
  {Grinstein}},\ }\href {https://doi.org/10.1088/0305-4470/13/2/024} {\bibfield
   {journal} {\bibinfo  {journal} {Journal of Physics A: Mathematical and
  General}\ }\textbf {\bibinfo {volume} {13}},\ \bibinfo {pages} {585}
  (\bibinfo {year} {1980})}\BibitemShut {NoStop}%
\bibitem [{\citenamefont {Ginsparg}(1988)}]{GINSPARG1988153}%
  \BibitemOpen
  \bibfield  {author} {\bibinfo {author} {\bibfnamefont {P.}~\bibnamefont
  {Ginsparg}},\ }\href
  {https://doi.org/https://doi.org/10.1016/0550-3213(88)90249-0} {\bibfield
  {journal} {\bibinfo  {journal} {Nuclear Physics B}\ }\textbf {\bibinfo
  {volume} {295}},\ \bibinfo {pages} {153} (\bibinfo {year}
  {1988})}\BibitemShut {NoStop}%
\bibitem [{\citenamefont {Kadanoff}\ and\ \citenamefont
  {Brown}(1979)}]{KADANOFF1979318}%
  \BibitemOpen
  \bibfield  {author} {\bibinfo {author} {\bibfnamefont {L.~P.}\ \bibnamefont
  {Kadanoff}}\ and\ \bibinfo {author} {\bibfnamefont {A.~C.}\ \bibnamefont
  {Brown}},\ }\href
  {https://doi.org/https://doi.org/10.1016/0003-4916(79)90100-3} {\bibfield
  {journal} {\bibinfo  {journal} {Annals of Physics}\ }\textbf {\bibinfo
  {volume} {121}},\ \bibinfo {pages} {318} (\bibinfo {year}
  {1979})}\BibitemShut {NoStop}%
\bibitem [{\citenamefont {Nomura}\ and\ \citenamefont
  {Kitazawa}(1996)}]{NomuraKitazawa1996}%
  \BibitemOpen
  \bibfield  {author} {\bibinfo {author} {\bibfnamefont {K.}~\bibnamefont
  {Nomura}}\ and\ \bibinfo {author} {\bibfnamefont {A.}~\bibnamefont
  {Kitazawa}},\ }\href@noop {} {\bibfield  {journal} {\bibinfo  {journal}
  {arXiv}\ } (\bibinfo {year} {1996})}\BibitemShut {NoStop}%
\bibitem [{\citenamefont {Cardy}(1986)}]{Cardy:1986aa}%
  \BibitemOpen
  \bibfield  {author} {\bibinfo {author} {\bibfnamefont {J.~L.}\ \bibnamefont
  {Cardy}},\ }\href
  {https://doi.org/https://doi.org/10.1016/0550-3213(86)90552-3} {\bibfield
  {journal} {\bibinfo  {journal} {Nuclear Physics B}\ }\textbf {\bibinfo
  {volume} {270}},\ \bibinfo {pages} {186} (\bibinfo {year}
  {1986})}\BibitemShut {NoStop}%
\bibitem [{\citenamefont {Hirata}(1999)}]{hirata1999}%
  \BibitemOpen
  \bibfield  {author} {\bibinfo {author} {\bibfnamefont {S.}~\bibnamefont
  {Hirata}},\ }\href {https://arxiv.org/abs/cond-mat/9912066} {\bibinfo {title}
  {Magnetization jump in the spin-1/2 xxz chain with next-nearest-neighbour
  coupling}} (\bibinfo {year} {1999}),\ \Eprint
  {https://arxiv.org/abs/cond-mat/9912066} {arXiv:cond-mat/9912066
  [cond-mat.str-el]} \BibitemShut {NoStop}%
\end{thebibliography}%

\end{document}